\documentclass[twocolumn,showpacs,amsmath,amssymb,prl,nofootinbib]{revtex4}
\usepackage{graphicx}
\usepackage{dcolumn}
\usepackage{bm}
\usepackage{psfig}

\newcommand{\ppbar}{p \overline{p}}
\newcommand{\pbar}{\overline{p}}
\newcommand{\Lambar}{\overline{\Lambda}}
\newcommand{\etap}{\eta^{\prime}}

\newcommand{\jpsi}{J/\psi}
\newcommand{\pip}{\pi^+}
\newcommand{\pin}{\pi^-}
\newcommand{\pio}{\pi^0}

\newcommand{\kk}{K^+K^-}

\newcommand{\g}{\gamma}

\newcommand{\ar}{\rightarrow}

\def\Journal#1&#2&#3(#4){#1{\bf #2}, #3 (#4)}

\def\bec{\begin{center}}
\def\eec{\end{center}}

\begin{document}

\title{ \bf \boldmath Observation of a broad $1^{--}$ resonant structure
around 1.5 GeV/$c^2$ in the $\kk$ mass spectrum in $J/\psi\to K^+K^-\pi^0 $}

\author{M.~Ablikim$^{1}$, J.~Z.~Bai$^{1}$, Y.~Ban$^{11}$,
X.~Cai$^{1}$, H.~F.~Chen$^{16}$, H.~S.~Chen$^{1}$, H.~X.~Chen$^{1}$, 
J.~C.~Chen$^{1}$, Jin~Chen$^{1}$, Y.~B.~Chen$^{1}$
% ...
\vspace{0.2cm} \\(BES Collaboration)\\
\vspace{0.2cm}
{\it $^{1}$ Institute of High Energy Physics, Beijing 100049, People's Republic of China\\
% ...
$^{11}$ Peking University, Beijing 100871, People's Republic of China\\
% ...
$^{16}$ University of Science and Technology of China, Hefei 230026, People's Republic of China\\
% ...
}}

\date{\today}

\begin{abstract}
{
A broad peak is observed at low $\kk$ invariant mass in
$J/\psi \rightarrow K^+K^-\pi^0$ decays found
in a sample of $5.8\times 10^7$ $\jpsi$ events collected with the BESII
detector.   
A partial wave analysis shows that the $J^{pc}$ of this structure
is $1^{--}$. Its pole position is determined to be
$(1576^{+49}_{-55}$(stat)$^{+98}_{-91}$(syst)) MeV/$c^2$
 - $i (409^{+11}_{-12}$(stat)$^{+32}_{-67}$(syst)) MeV/$c^2$.
These parameters are not compatible with any known meson resonances.
}

\end{abstract}

\maketitle


The $\jpsi$ meson has been useful for searches for new hadrons and studies of 
light hadron spectroscopy. Recently, a number of new structures have been 
observed in $\jpsi$ decays. These include strong near-threshold mass 
enhancements in the $p\pbar$ invariant mass spectrum from 
$\jpsi\rightarrow\gamma p\pbar$ decays~\cite{gpp}, the $p\Lambar$ and the 
$K^-\Lambar$ mass spectra in  $J/\psi \rightarrow p K^- \Lambar$ 
decays~\cite{pkl}, the $\omega\phi$ mass spectrum in the double-OZI suppressed 
decay $\jpsi\ar\g\omega\phi$~\cite{goph}, and a new resonance, the $X(1835)$, 
in $\jpsi\ar\g\pip\pin\etap$ decays~\cite{x1835}. Some of these new structures 
have not been observed in other experiments.  For example, the strong $\ppbar$
mass threshold enhancement is neither observed in $\ppbar$ cross section 
measurements, nor in B decays~\cite{ichep04}. These experimental observations 
are unexpected and have stimulated interest in searching for other new hadron 
states in $\jpsi$ decays. Since the $\jpsi$ has $J^{PC}=1^{--}$ and zero
isospin, its decays are particularly useful for spin-parity and isospin 
determinations of hadronic states found in its decays. In this Letter, we 
report the first observation of a broad $1^{--}$ resonant structure in the 
invariant mass spectrum of $\kk$ in the channel 
$J/\psi \rightarrow K^+K^-\pi^0$.  The results come from an analysis of $5.8
\times 10^7$ $J/\psi$ decays detected with the upgraded Beijing Spectrometer 
(BESII) at the Beijing Electron-Positron Collider (BEPC).


BESII is a large solid-angle magnetic spectrometer that is described in detail
in Ref.~\cite{BESII}. Charged particle momenta are determined with a resolution
of $\sigma_p/p = 1.78\%\sqrt{1+p^2(\mbox{\rm GeV/c}^2)}$ in a 40-layer
cylindrical main drift chamber (MDC). Particle identification is accomplished
by specific ionization ($dE/dx$) measurements in the MDC and time-of-flight
(TOF) measurements in a barrel-like array of 48 scintillation counters. The
$dE/dx$ resolution is $\sigma_{dE/dx} = 8.0\%$; the TOF resolution is measured
to be $\sigma_{TOF} = 180$~ps for Bhabha events. Outside of the time-of-flight
counters is a 12-radiation-length barrel shower counter (BSC) comprised of gas
tubes interleaved with lead sheets. The BSC measures the energies and
directions of photons with resolutions of
$\sigma_E/E\simeq 21\%/\sqrt{E(\mbox{GeV})}$, $\sigma_{\phi} = 7.9$ mrad,
and $\sigma_{z}$ = 2.3 cm. The iron flux return of the magnet is instrumented
with three double layers of counters that are used to identify muons. In this
analysis, a GEANT3-based Monte Carlo (MC) package with detailed consideration
of the detector performance is used. The consistency between data and MC has
been carefully checked in many high-purity physics channels, and the agreement
is reasonable~\cite{simbes}.

\begin{figure}[hbtp]
          \centerline{\psfig{file=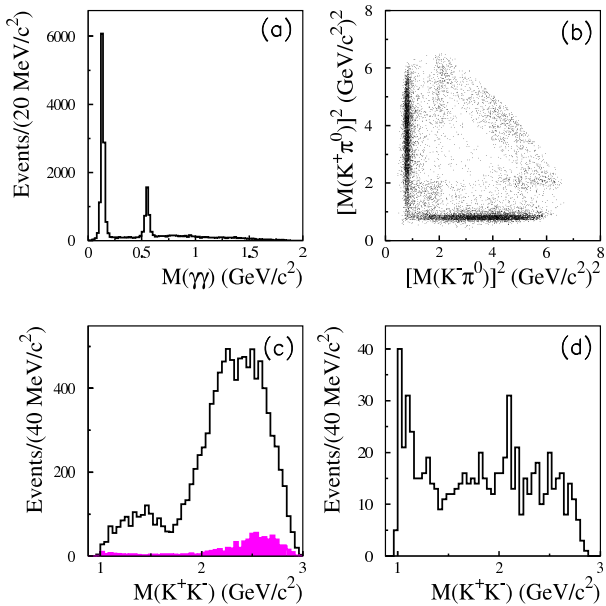,height=8.0cm,width=8.0cm}}
          \caption{
           (a) The $\g\g$ invariant mass distribution.
           (b) The Dalitz plot for $\kk\pio$ candidate events.
           (c) The $\kk$ invariant mass distribution for $\kk\pio$
               candidate events;
               the solid histogram is data and the shaded histogram
               is the background (normalized to data).
           (d)  The $\kk$ invariant mass distribution for the $\pio$ mass
                sideband events (not normalized).}
    \label{data}
    \end{figure}

Candidate $J/\psi \to K^+K^-\pi^0$ events are required to have two oppositely 
charged tracks, each of which is well fitted to a helix that is within the
polar angle region $|\cos \theta|<0.8$ and with a transverse momentum larger 
than $70$~MeV/c. For each track, the TOF and $dE/dx$ information are combined 
to form particle identification confidence levels for the $\pi, K$ and $p$ 
hypotheses; the particle type with the highest confidence level is assigned to
each track. The two charged tracks are required to be identified as kaons. 
Candidate photons are required to have an energy deposit in the BSC greater 
than 50~MeV and to be isolated from charged tracks by more than $15^{\circ}$; 
at least two photons are required. A four-constraint (4C) energy-momentum 
conservation kinematic fit is performed to the $K^+K^-\gamma\gamma$ hypothesis
and $\chi^2 < 10$ is required.  For events with more than two selected photons,
the combination with the smallest $\chi^2$ is chosen. Figure~\ref{data}(a) 
shows the fitted $\gamma\gamma$ invariant mass distribution, where a $\pi^0$ 
signal is evident. Candidate $\pi^0$s are identified by the requirement $\mid
M_{\gamma\gamma}-m_{\pi^0} \mid<0.04$ GeV/$c^2$. To reduce the background 
events with mis-reconstructed $\pio$'s, the energies ($E_{\g 1}, E_{\g 2}$) of 
the two photons from the $\pio$ are required to satisfy 
$|(E_{\g 1} - E_{\g 2})/(E_{\g 1} + E_{\g 2})| < 0.8$. To suppress background 
from the radiative decay process $J/\psi \to \gamma\pi^0K^+K^-$,  we require 
the candidate events to fail a five-constraint kinematic fit to the 
$\gamma\pi^0K^+K^-$ hypothesis ( $\chi^2_{\gamma \pi^{0} \kk}>50$ ), where the 
invariant mass of the $\g\g$ pair associated with the $\pio$ is constrained to 
$m_{\pio}$~\cite{optimum}.

The Dalitz plot for the selected events is shown in Fig.~\ref{data}(b),
where a broad $K^+K^-$ band is evident in addition to the $K^*(892)$ 
and $K^*(1410)$ signals.  This band corresponds to the 
broad peak observed around 1.5 GeV/$c^2$ in the $K^+K^-$ invariant mass
projection shown in Fig.~\ref{data}(c).

Backgrounds for this decay channel have been studied using both data
and MC.  The cleanliness of the $\pio$ signal shown in Fig. \ref{data}(a) 
indicates that non-$\pio$ background processes correspond to only about 
$2\%$ of the selected events.  The $\kk$ mass distribution for the $\pio$ 
sideband events, shown in Fig.~\ref{data}(d), has a different character from 
that of the signal (shown in Fig. \ref{data}(c)). A Monte Carlo (MC) study 
indicates that background from $\jpsi\ar\rho\pi\ar\pip\pin\pio$ decays, which 
produces a $\pio$ peak, comprises about $6\%$ of the selected event sample.  
These events are dominantly peaked at high $\kk$ masses as shown by the shaded 
histogram in Fig. \ref{data}(c);  the $\pip\pin\pio$ contamination to the low 
$\kk$ invariant mass region is almost completely eliminated by the particle 
identification and kinematical fit requirements. Backgrounds from processes 
such as $\jpsi\ar\omega\pip\pin$ and $\jpsi\ar\g\eta_C\ar\g\kk\pio$ are found 
to be negligible. From these studies, we conclude that the broad low $\kk$ mass
peak is not from any background process.

A partial wave analysis (PWA) is used to determine the mass, width and
spin-parity of the broad peak at low mass, which is denoted as $X$.
The amplitudes are constructed using the relativistic covariant
tensor amplitude method~\cite{pwa-method}, and the maximum likelihood method 
is used in the fit. The decay process is modeled by a phase space 
contribution (i.e., direct three body decays with correct angular 
distributions) plus several sequential two-body decays: $J/\psi\to X\pi^0 $,
$X \to K^+K^-$,  $J/\psi\to \rho\pi^0 $, $\rho \to \kk$ and 
$J/\psi\to (K^*)^{\pm}K^{\mp} $, 
$(K^*)^{\pm} \to K^{\pm}\pi^0 $. The broad resonance $X$ is parameterized by 
a Breit-Wigner (BW) function with a mass-dependent width~\cite{gam-s}.  
Background contributions are removed by subtracting the log-likelihood value 
of background events from that of data~\cite{bes-pwa}.

\begin{figure}[hbtp]
          \centerline{\psfig{file=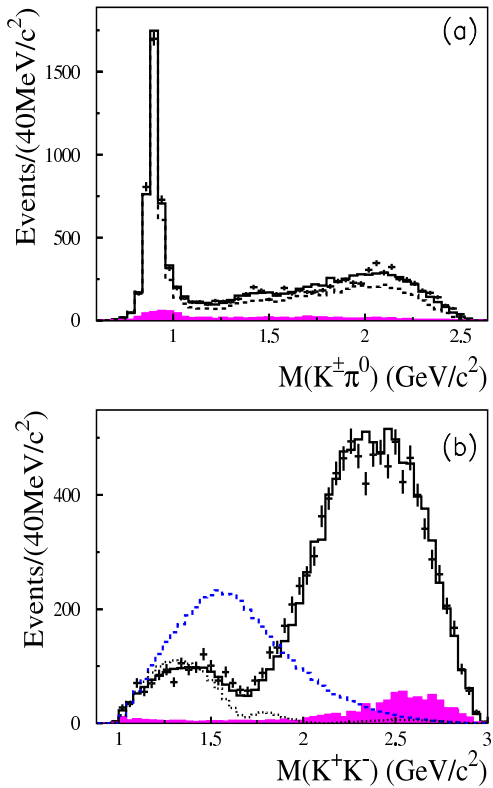,height=10.5cm,width=6.5cm}}
          \caption{
           (a) The $K^\pm\pio$ invariant mass distribution;
               the error bars are data, the solid histogram is the PWA fit
               projection, the dashed histogram is the $1^-$ component
               of $K^\pm\pio$ system and the shaded histogram is the 
               background.
           (b) The $\kk$ invariant mass distribution;
               the error bars are data, the solid histogram is the PWA fit
               projection, the upper dashed histogram is the X component, 
               the lower dotted histogram is the $1^{--}$ component
               of $\kk$ system and the shaded histogram is the background.}
    \label{mass}
    \end{figure}

\begin{figure}[hbtp]
          \centerline{\psfig{file=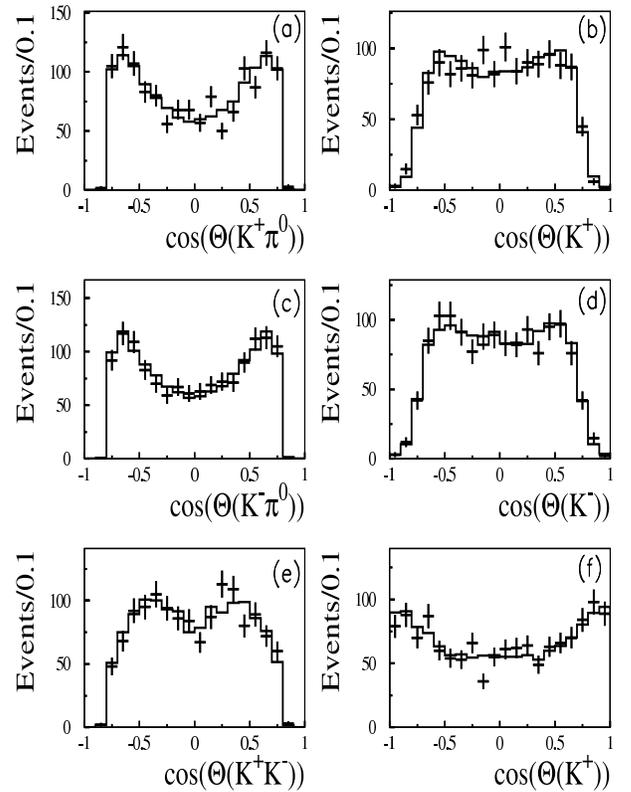,height=10.5cm,width=8.0cm}}
          \caption{ Angular distributions of the events with $M_{\kk} < 1.7$
           GeV/$c^2$; error bars are data and solid histograms are the PWA
           fit projections:
           (a), (c) and (e) are the angular distributions of $K^+\pio$,
            $K^-\pio$ and $\kk$ system in the laboratory frame;
           (b), (d) and (f) are the angular distributions of $K^+$, $K^-$
           and $K^+$ in the center of mass frames of the $K^+\pio$, $K^-\pio$
           and $\kk$ systems, respectively.}
    \label{cos}
    \end{figure}

Five components, the $X$, $K^\star(892)$, $K^\star(1410)$, $\rho(1700)$ 
and phase space, are included in the PWA fit. The $K^\star(892)$,
$K^\star(1410)$ and $\rho(1700)$ parameters are fixed at Particle Data Group 
(PDG) values~\cite{pdg}, and their uncertainties are included in the 
systematic errors on the parameters of the $X$. Parity conservation in 
$\jpsi\to \kk\pio$ decay restricts the possible spin-parity of the $\kk$ system
to $1^{--}, 3^{--}, ...$. The PWA determines the spin-parity of $X$ to be 
$1^{--}$. The log-likelihood value of the fit becomes worse by 325 for a 
$J^{PC}$ assignment of $3^{--}$; even higher spin states are unlikely at such 
a low mass. The $\sigma$ and $\kappa$ resonance studies at BESII~\cite{bes-pwa}
show that the parameters of a broad resonance are better described by the pole 
position since it has less dependence on the details of the BW formula that is 
used. From the PWA fit, the $X$ pole position is determined to be 
$(1576^{+49}_{-55})$ MeV/$c^2$ - $i (409^{+11}_{-12})$ MeV/$c^2$, and the 
branching ratio is $B(\jpsi\ar X \pi^0)\cdot B(X\ar \kk)$ =
$(8.5\pm0.6)\times 10^{-4}$, where the errors are statistical only. 
In the PWA fit, there is large destructive interference between the $X$, the 
$\rho(1700)$ and phase space, which produces the 'hole' seen in the center of 
the Dalitz plot. The comparisons of the mass distributions between the data and
the PWA fit projections (weighted by MC efficiencies) are displayed in 
Fig.~\ref{mass}. The angular distributions of the events with 
$M_{\kk} < 1.7$ GeV/$c^2$ are shown in Fig.~\ref{cos}.

Each of the five PWA components has a statistical significance
that is larger than $5\sigma$. In the PWA fit, when we remove the 
$X$, $K^\star(892)$, $K^\star(1410)$, $\rho(1700)$ and phase space components 
one at a time,  the log-likelihood values worsen by 533, 11438, 465, 28 and 
130, respectively. If we replace the $X$ by three additional interfering 
resonances, the $\rho(770)$, the $\rho(1900)$ and the $\rho(2150)$, the 
log-likelihood value worsens by 85. The  broad resonant structure $X$ is 
unlikely to be due to the $\rho(1450)$; in addition to the fact that the 
parameters of the $X$ resonance are incompatible with those of $\rho(1450)$, 
including various systematic uncertainties studied below, the $\rho(1450)$ is 
known to have a very small branching fraction to $\kk$ ($<1.6\times 10^{-3}$ 
at 95 $\%$ C.L.)~\cite{pdg}. We conclude that the broad peak at low $\kk$ mass 
is not described by any known mesons or their interferences.

If we do not include a $\rho(1700)$, which has the lowest relative statistical 
significance ($7.2\sigma$) of the five components used in the final PWA 
fit~\cite{rho1700}, the pole position moves to  $(1428^{+17}_{-18}$(stat)) 
MeV/$c^2$ - $i (536^{+15}_{-12}$(stat)) MeV/$c^2$, and the branching ratio
becomes  $B(\jpsi\ar X \pi^0)\cdot B(X\ar \kk)$ =
$(6.3\pm0.6$(stat))$\times 10^{-4}$.

We have studied the systematic uncertainties from inclusion of other possible
resonances ($\rho(770)$, $\rho(1900)$, $\rho(2150)$, $K^*_2(1430)$, 
$K^*(1680)$ and the possible $K^*(2075)$ that is indicated by the $p\Lambar$ 
mass threshold enhancement~\cite{gpp}) in the PWA fits~\cite{lowsig},  
and the use of different BW formulae,  background levels, parameters of the 
$K^\star(892)$, $K^\star(1410)$ and $\rho(1700)$, MDC wire resolution 
simulation, particle identification, photon selection and the total number of 
$\jpsi$ events~\cite{fangss}. We find that the inclusion of other 
resonances causes the dominant shifts in the parameters of the $X$.
For example, if the $\rho(770)$ is included in the PWA fit, the log-likelihood
value improves by 13, and the pole position and the branching ratio
change by $(-58 - i (-56))$ MeV/$c^2$ and -30$\%$, respectively. These changes 
are considered as systematic uncertainties. The total systematic uncertainties
on the pole position and branching ratios are $(^{+98}_{-91}$ - 
$i ^{+32}_{-67})$ MeV/$c^2$ and $^{+31\%}_{-42\%}$,  respectively.


In summary, we observe a broad peak at low $\kk$ invariant mass in the
channel $J/\psi \rightarrow K^+K^-\pi^0$. A partial wave analysis
shows that the $J^{pc}$ of this structure
is $1^{--}$. Its pole position is determined to be
$(1576^{+49}_{-55}$$^{+98}_{-91})$ MeV/$c^2$
 - $i (409^{+11}_{-12}$$^{+32}_{-67})$ MeV/$c^2$,
and the branching ratio is $B(\jpsi\ar X \pi^0)\cdot B(X\ar \kk)$=
$(8.5\pm0.6^{+2.7}_{-3.6})\times 10^{-4} $, where the first errors are 
statistical and the second are systematic.  These
parameters are not compatible with any known meson resonances~\cite{pdg}.

To understand the nature of the broad $1^{--}$ peak, it is important to search
for a similar structure in $\jpsi\ar K_S K^\pm \pi^\mp$ decays to determine
its isospin. It is also intriguing to search for $K^*K, KK\pi$ decay modes.
In the mass region of the $X$, there are several other $1^{--}$ states, such as
the $\rho(1450)$ and $\rho(1700)$, but the width of the $X$ is much
broader than the widths of these other mesons. This may be an
indication that the
$X$ has a different nature than these other mesons.  For example,
very broad  widths are expected for multiquark states~\cite{jpsi}.

The BES collaboration acknowledges the staff of BEPC for the excellent
performance of the machine. This work is supported in part by the National
Natural Science Foundation of China under contracts Nos. 10491300, 10225524,
10225525, 10425523, 10521003, the Chinese Academy of Sciences under contract
No. KJ 95T-03, the 100 Talents Program of CAS under Contract Nos. U-11, U-24,
U-25, and the Knowledge Innovation Project of CAS under Contract Nos.
KJCX2-SW-N10, U-602, U-34 (IHEP); by the National Natural Science Foundation of
China under Contract No. 10175060 (USTC); and by the Department of Energy under
Contract No. DE-FG03-94ER40833 (U Hawaii).


\begin {thebibliography}{99}
\bibitem{optimum}
The selection criteria are optimized by maximizing
$N_s/\sqrt{N_{tot}}$, where $N_s$ is the expected number of signal events and
$N_{tot}$ is the total expected number of signal and background events.
\bibitem{rho1700} Because data-MC inconsistencies are not included in the 
    fit, the significance of the $\rho(1700)$ may be not high enough to 
    conclude that its inclusion in the fit is necessary.  We therefore list 
    the results of the fit without $\rho(1700)$.
\bibitem{lowsig} We include the resonances with sigificances below
    $5\sigma$ as systematic uncertaities.
\end{thebibliography}
\end{document}